\documentclass[10pt]{wlscirep}
\usepackage[utf8]{inputenc}
\usepackage[T1]{fontenc}
\usepackage{graphicx}%
\usepackage{multirow}%
\usepackage{amsmath,amssymb,amsfonts}%
\usepackage{amsthm}%
\usepackage{mathrsfs}%
\usepackage[title]{appendix}%
\usepackage{xcolor}%
\usepackage{textcomp}%
\usepackage{manyfoot}%
\usepackage{booktabs}%
\usepackage{algorithm}%
\usepackage{algorithmicx}%
\usepackage{algpseudocode}%
\usepackage{listings}%
\usepackage{graphicx}
\usepackage{subcaption}
\usepackage[numbers]{natbib}
\usepackage{booktabs}
\usepackage{multirow}
\usepackage{amsmath}
\usepackage{tikz}
\usepackage{yquant}
\usepackage{xcolor}
\usepackage{pgfplots}
\pgfplotsset{compat=newest}
\usetikzlibrary{quantikz}
\usetikzlibrary{positioning}

\definecolor{darkred}{RGB}{200,0,0}
\definecolor{darkgreen}{RGB}{0,130,0}

\newcommand{\rev}[1]{}
\renewcommand{\rev}[1]{{\color{black} {#1}}}  

\newcommand{\revv}[1]{}
\renewcommand{\revv}[1]{{\color{black} {#1}}}  

\title{Quantum Global Minimum Finder based on Variational Quantum Search}

\author[1,+]{Mohammadreza Soltaninia}
\author[1,*+]{Junpeng Zhan}
\affil[1]{Department of Electrical Engineering, Alfred University, Alfred, NY 14802, USA}

\affil[*]{zhanjunpeng@gmail.com}

\affil[+]{these authors contributed equally to this work}

\begin{abstract}
The search for global minima is a critical challenge across multiple fields including engineering, finance, and artificial intelligence, particularly with non-convex functions that feature multiple local optima, complicating optimization efforts. We introduce the Quantum Global Minimum Finder (QGMF), an innovative quantum computing approach that efficiently identifies global minima. QGMF combines binary search techniques to shift the objective function to a suitable position and then employs Variational Quantum Search to precisely locate the global minimum within this targeted subspace. 
\rev{Designed with a $O(n)$-depth circuit architecture, QGMF also utilize the logarithmic benefits of binary search to enhance scalability and efficiency. }
This work demonstrates the impact of QGMF in advancing the capabilities of quantum computing to overcome complex non-convex optimization challenges effectively. 
\end{abstract}
\begin{document}

\flushbottom
\maketitle

\thispagestyle{empty}

\section{Introduction}

\label{sec:intro}
Optimization challenges permeate a diverse range of fields, including science \cite{opt_in_sience}, engineering \cite{boyd2004convex}, and economics \cite{opt_in_economics}. Central to these challenges is the quest to find the global minimum of a function, representing the most optimized state or solution within a given domain. Despite considerable research efforts \cite{ge2015escaping, choromanska2015loss}, accurately pinpointing global minimum in complex landscapes, characterized by multiple local minima, remains challenging.
Traditional strategies, such as the brute-force approaches mentioned in \cite{mahoor2017hierarchical, ploskas2022review}, necessitate traversing all elements in the search space, which becomes impractical for large datasets due to the "curse of dimensionality".

Alternative optimization techniques, including gradient-based \cite{gradiant_desc}, non-gradient-based \cite{grad_free}, and genetic algorithms \cite{genetic_review}, offer strategies to navigate the search space more efficiently. Yet, these methods often fall short in guaranteeing the discovery of the global optimum, particularly in complex, non-convex functions \cite{clarke1990optimization} that pose the risk of trapping these algorithms in local optima.

Quantum computing emerges as a revolutionary paradigm, offering a novel approach to these longstanding optimization challenges. Unlike classical systems, which process information in binary bits, quantum systems utilize qubits, enabling the representation of multiple states simultaneously through superposition \cite{knill2010quantum}. This capability allows quantum algorithms to explore vast search spaces more efficiently, potentially accelerating the discovery of global minimum for specific problem classes.

Among the quantum computing methodologies, Quantum Annealing (QA) has gained prominence for its potential in solving a subset of optimization problems known as "combinatorial optimization problems'' \cite{qan}. QA leverages the principles of quantum mechanics to navigate the search space, aiming to converge on the global minimum by exploiting quantum tunneling and superposition. However, the practical application of QA is constrained by the need for adiabatic evolution \cite{QALRevModPhys.90.015002} and faces challenges in guaranteeing solution optimality within realistic computational timeframes.

Grover's algorithm \cite{grover1996fast} is a quantum algorithm that provides a quadratic speedup for searching unsorted databases, finding a specific item in $O(\sqrt{N})$ time, where $N$ is the number of items, compared to $O(N)$ time in classical computing. In \cite{gfp1}, the author modified the Grover's quantum algorithm \cite{grover1996fast} to solve real-world problems in identifying a global minimum, employing a relatively small number of quantum bits to find the global minimum. Despite the reduction to $O(\sqrt{N})$, the number of function evaluations still exhibits an exponential dependency on the number of qubits, given that $N = 2^{n}$, where $n$ is the number of qubits.

Although Grover's algorithm and the quantum amplitude amplification technique \cite{amplitude} provide quadratic speedup across various scientific and computational problems, their practical application faces limitations due to the exponential growth in quantum circuit complexity as more qubits are added. The advent of Variational Quantum Search (VQS) \cite{zhan2023variational} introduces an innovative method that capitalizes on variational quantum algorithms and parameterized quantum circuits. VQS notably improves the odds of identifying desirable elements, while ensuring the maximal depth of quantum circuits increases linearly with the qubit count, a fact validated for systems up to 26 qubits. 


This paper introduces the Quantum Global Minimum Finder (QGMF), an innovative quantum algorithm designed to efficiently locate the global minima of any given function, which is challenging due to potential multiple local optima. The QGMF algorithm combines a binary search technique with the VQS \rev{(the VQS part does not have an oracle)}. Characterized by its \rev{$O(n)$-depth} circuit design, \rev{the QGMF also leverage} the logarithmic advantage of binary search to enhance scalability and efficiency. This method represents an advance in tackling complex optimization problems across various fields by harnessing the power of quantum computing.

\revv{In \cite{Barito2005} and \cite{Gilliam2021groveradaptive}, the adaptive search method involves iteratively applying Grover's algorithm to the improving region -- defined as the set of points where the objective function value is better than a current threshold -- to find better solutions. In contrast, the QGMF proposed in this paper employs a binary search to efficiently determine a shift value that adjusts the entire objective function up or down. This shifted objective function has a limited number of negative objective values, allowing the VQS to effectively identify the minimum among them. This approach not only achieves logarithmic efficiency due to the binary search but also benefits from the reduced circuit depth of VQS compared to Grover's algorithm, enhancing practicality for near-term quantum devices.
}

The remainder of the paper is structured as follows: In Section \ref{sec:method}, we detail the QGMF, including its components such as quantum circuit, binary search, and VQS. Section \ref{sec:tests} presents a complexity analysis and case study to evaluate the performance of the QGMF. Finally, Section \ref{sec:conc} explores the broader implications of our findings and suggests directions for future research, while considering the limitations of the current study.

\section{Method: QGMF} 
This section details the QGMF, beginning with a high-level description to introduce the general concept and operational framework of the QGMF. It describes each component of the QGMF, including the quantum circuits used for an Oracle, a quantum adder, and the VQS \rev{without an oracle}. Additionally, the method incorporates a binary search technique to efficiently shift the objective function such that it helps VQS to effectively find the global minimum. The section also includes the pseudocode for the QGMF algorithm (Algorithm~\ref{alg:binary_vqs}) and visual representations of the quantum circuits involved (Figure~\ref{fig:vqs_binary}), offering a comprehensive guide to understanding and implementing the QGMF. 

\label{sec:method}
\subsection{High-Level Description of QGMF}

\paragraph{Assumption:} We consider an arbitrary function $f(x)$ without any restrictions on its properties, including convexity, continuity, differentiability, discreteness, etc.

\paragraph{Problem:} Our goal is to determine the global minimum ($G_{m}$) of the function $f(x)$ over a specific domain $D$.

\paragraph{Solution:}
 The idea is to determine $G_{m}$ through an iterative process, as depicted in Figure \ref{fig:high-level-graph}. In this approach, we shift the function $f(x)$ along the y-axis by an amount of $s_i$ in the $i$th iteration. This process continues until the number of negative values of $f(x)$ falls between 1 and a predefined threshold $T$, with the count of negative values determined by the VQS. 
 Then we find the minimum of existing negative values ($M_r$) and utilize Eq. (\ref{eq:gmin}) to compute the global minimum. 
\begin{equation}
	\label{eq:gmin}
	G_{m} = M_r - s_i
\end{equation}

It is important to note that the function $f(x)$ may 
have an infinite number of values. 
Nevertheless, given that $f(x)$ is encoded through sampling, the count of evaluated $f(x)$ values becomes finite.

\begin{figure}[H]
	\centering
	\begin{tikzpicture}

		\begin{axis}[
			axis lines = middle,
			xlabel = $x$,
			ylabel = {$y$},
			xticklabels={,,}, 
			yticklabels={,,}, 
			tick style={draw=none}, 
			width=12cm, 
			height=8cm, 
			enlargelimits=true, 
			enlarge x limits={rel=0.1}, 
			enlarge y limits={upper,rel=0.1}, 
			legend style={at={(1.05,1)}, anchor=north west}, 
			]
			
			\addplot [
			domain=-0:10, 
			samples=80, 
			color=black,
			dashed,
			]
			{sin(2.5*deg(x)) + cos(0.7*deg(x))+7} 
			node[pos=0.85, above, font=\small] {$f(x)$};
			
			
			\addplot [
			domain=0:10, 
			samples=80, 
			dashed,
			]
			{sin(2.5*deg(x)) + cos(0.7*deg(x))+5-6.5} 
			node[pos=0.85, above, font=\small] {$f(x)+s_1$};
			
			\addplot [
			domain=0:10, 
			samples=80, 
			dashed,
			mark options={scale=0.9} 
			]
			{sin(2.5*deg(x)) + cos(0.7*deg(x))+5-3.5}
			node[pos=0.85, above, font=\small] {$f(x)+s_2$};
			
			\node[label={180:{$G_m$}},circle,fill,color=blue,inner sep=1.5pt] at (axis cs:4.41,5) {};
			\node[label={[text=black]272:{$G_m+s_2$}},circle,fill,color=blue,inner sep=0pt] at (axis cs:4.3,-0.5) {};
         \node[circle,fill,color=blue,inner sep=1.5pt] at (axis cs:4.4,-0.52) {};

			\node[label={180:{$G_m + s_1$}},circle,fill,color=blue,inner sep=1.5pt] at (axis cs:4.41,-3.5) {};
			
		\end{axis}
	\end{tikzpicture}
	
	\caption{High-level overview of the QGMF algorithm for finding the global minimum of $f(x)$. In this specific example, since $f(x)$ is initially positive for all inputs, it needs to be shifted downward along the y-axis to generate negative values. This downward shift is accomplished by shifting the function by $s_1$. After the initial downward shift by $s_1$, the function produces a significant number of negative values (exceeding our specified threshold $T$), indicating the need for a smaller shift by $s_2$. Following this adjustment, the count of negative values of $f(x)+s_2$ becomes smaller than $T$, enabling us to easily identify the minimum among the remaining negative numbers denoted as $M_r$. Finally, we utilize Eq. (\ref{eq:gmin}) to calculate the value of $G_m$.}
	\label{fig:high-level-graph}
\end{figure}

Conforming to the "quantum Church-Turing thesis" \cite{vaz:doi:10.1137/S0097539796300921}, it is conjectured that a quantum computer, given sufficient resources, can effectively simulate any classical computation and, in certain scenarios, execute it more efficiently than classical counterparts. Consequently, we can realize any classical function $f(x)$ as a quantum oracle $\hat{O}_f$, enabling the exploitation of quantum computing's distinctive properties within QGMF.

To identify the global minimum of a classical function, it is essential to transform this function into a quantum oracle $\hat{O}_f$. This transformation has been the focus of various studies, such as those documented in \cite{func1}, which explore methodologies for representing classical functions within quantum frameworks. Building upon this foundation, quantum arithmetic operations—highlighted in both \cite{arithm} and our previous work \cite{zhan2023quantum}—introduce the capabilities for executing multiplication, addition, and averaging within a quantum domain. However, the efficient conversion of real-valued functions from $f(x)$ to $\hat{O}_f|x\rangle$ demands further exploration. Past research, including efforts by \cite{seidel2022efficient}, has begun to tackle the complexities associated with quantum arithmetic for real-valued functions. Integrating these established techniques into our approach promises a robust pathway towards achieving our objective.





\makeatletter
\DeclareRobustCommand\rvdots{%
	\vbox{%
		\baselineskip4\p@\lineskiplimit\z@%
		\kern-\p@%
		\hbox{.}\hbox{.}\hbox{.}%
	}%
}


\subsection{Defining the Oracle}
To exploit quantum parallelism, we first apply Hadamard gates to create a uniform superposition of different inputs (Eq. (\ref{eq:phi0})). This prepares a quantum state that encompasses all inputs simultaneously. 

\rev{
\begin{equation}
	\label{eq:phi0}
	|\phi_0\rangle = |0\rangle^{\otimes m-1} \otimes H^{\otimes n} |0\rangle^{\otimes n} = \frac{1}{2^{\frac{n}{2}}} \sum_{x=0}^{2^{n-1}} |0\rangle^{\otimes {m-1}}|x\rangle
\end{equation}
}

The next step in QGMF is to define the quantum oracle $\hat{O}_f$ which takes the input \rev{$|0,x\rangle$} and yields the output \rev{$|f(x),x\rangle$} in 2's complement format, as expressed in Eq.  (\ref{eq:oracle1}). Section \ref{sec:prereq} in Appendix details the methodologies for data embedding, focusing on basis embedding and 2's complement. 

\rev{
\begin{equation}
	\label{eq:oracle1}
	\hat{O}_f|0,x\rangle = |f(x),x\rangle
\end{equation}
}

The oracle is typically treated as a black-box quantum operator, with the objective of determining its global minimum while considering its output in 2's complement representation.

To prevent overflow during the shifting phase, an additional overflow qubit was introduced immediately following the oracle. Utilizing a CNOT gate in conjunction with the MSQ entangles it with the MSQ, ensuring adherence to the 2's complement conditions (e.g., sign extension) \cite{sign_binary} (refer to section \ref{sec:prereq}).

If needed, one can add any number of extra overflow qubits by following the same procedure. However, for our considerations, \rev{we assumed that $f(x) \subseteq [-2^{m-1},2^{m-1}-1]$. }
Additionally, given the worst-case scenario due to the lower and upper bounds of the shift $s_i$ in Algorithm~\ref{alg:binary_vqs}, where $s_i \leftarrow \lfloor(Low + High)/2\rfloor$, we have \rev{$s_i \subseteq [-2^{m-1},2^{m-1}-1]$}. Indeed, we need to accommodate the range \rev{ $[-2^{m}, 2^{m}-2]$}, requiring \rev{$m$} bits. Consequently, introducing one overflow qubit is sufficient.

			
			
			

By applying the oracle to the superposition of all potential inputs, we acquire a superposition of all oracle's outputs, as outlined in Eq. (\ref{eq:psi}), where $|\phi^{\prime}_0\rangle=\hat{O}_f|\phi_0\rangle$.   

\begin{equation}
	\label{eq:psi}
	|\phi_1\rangle = |0\oplus |\phi^{\prime}_0\rangle_{MSQ}\rangle \otimes |\phi^{\prime}_0\rangle
\end{equation}

\subsection{Quantum Adder} 
\label{subsec:shifting}

The next step is implementing the adder circuit ("$A(s_i)$") to add state $|\phi_{1}\rangle$ with the value $s_i$, which generates $|\phi_{2}(s_i)\rangle = |\phi_{1}+s_i\rangle$. This can be achieved by using the Quantum Fourier Transform (QFT)-based adder \cite{arithm}. Initially, we convert the state $|\phi_{1}\rangle$ from the computational basis to the Fourier basis by applying the QFT to it, resulting in a new state in the QFT basis denoted as $|\psi_1\rangle$ (see Eq. (\ref{eq:adder_1_prime})).

\begin{equation}
	\label{eq:adder_1_prime}
	|\psi_1\rangle = \mathrm{QFT} \left| {\phi_1}\right\rangle
\end{equation}

Subsequently, Eq. (\ref{eq:adder_1}) is employed to rotate the $j$-th qubit by an angle of $s_i\times\frac{\pi}{2^{j}}$ using the $R_Z$ gate. This rotation introduces a new phase of $(\phi_{1}+s_i)\times\frac{\pi}{2^{j}}$ for the $j$-th qubit where $0\leq j \leq n$. The resulting state after these rotations is denoted as $|\psi_2(s_i)\rangle$.

\begin{equation}
	\label{eq:adder_1}
	|\psi_2(s_i)\rangle = R_Z(s_i \cdot \frac{\pi}{2^0}) \left| \psi_1 \right\rangle_0 \otimes \ldots \otimes R_Z(s_i \cdot \frac{\pi}{2^{n}}) \left| \psi_1 \right\rangle_{n}
\end{equation}

Finally, we apply $QFT^{-1}$ to return to the computational basis, yielding $|\phi_2(s_i)\rangle$ as shown in Eq. (\ref{eq:adder_2}).

\begin{equation}
	\label{eq:adder_2}
	|\phi_2(s_i)\rangle=\mathrm{QFT}^{-1}|\psi_2(s_i)\rangle \stackrel{\text{set}}{=} |\phi_{1}+s_i\rangle
\end{equation}

It is essential to underscore that the operator $A(s_i)$ is designed to work exclusively with positive integers. However, in situations where the shift "$s_i$" is a negative integer, we need to perform a conversion from the 2's complement form of the negative number to its positive decimal representation and then apply it to the $A$ operator. For example, if we intend to use $A(-2)$ with a 3-qubit configuration, we first convert $-2$ to its 2's complement form, which is ${(110)}_2$, equivalent to $6$ in positive decimal notation. Consequently, we should utilize $A(6)$ in this scenario.

\begin{algorithm}
	\caption{\ \ QGMF algorithm for finding the $G_{m}$ of $\hat{O}_f|x\rangle$.}\label{alg:binary_vqs}
	\begin{algorithmic}[1]
		\State \textbf{Input:}
		\State \hspace{\algorithmicindent} $\hat{O}_f:$ Quantum oracle in 2's complement binary format
		\State \hspace{\algorithmicindent} $n:$ Number of qubits
		\State \hspace{\algorithmicindent} $T:$ Threshold $\geq 1$
		\State \textbf{Output:}
		\State \hspace{\algorithmicindent} $G_{m}:$ Global minimum of $\hat{O_f}|x\rangle$
		\State
		\State $Low \leftarrow -2^{n - 1}$
		\State $High \leftarrow 2^{n - 1}$
		\State $i \leftarrow 0$
		\While{True}
		\State $s_i \leftarrow \lfloor(Low+High)/2\rfloor$
		\State $\theta_{best} \leftarrow$ Run VQS on $|\phi_2(s_i)\rangle$
		\State $results \leftarrow$ Measure the state $|\phi_3(s_i,\theta_{\text{best}})\rangle$
		\State $N_{neg} \leftarrow$ Find the number of distinct negative values in $results$ \revv{(once the number of distinct negative values measured exceeds $T$, the measurement process is terminated)}
		\If{$N_{neg} \in [1,T]$}
		\State \textcolor{black}{ Denoted the minimum value among all $y_j$ as $y_{\text{min}}$ }
        \State \textcolor{black}{ $s_i \leftarrow s_i+|y_{\text{min}}| + \epsilon $ }
		\State \textcolor{black}{ $\theta_{best} \leftarrow$ Run VQS on $|\phi_2(s_i)\rangle$ }
		\State \textcolor{black}{ $results \leftarrow$ Measure the state $|\phi_3(s_i,\theta_{\text{best}})\rangle$ }
		\State \textcolor{black}{ $N_{neg} \leftarrow$ Find the number of distinct negative values in $results$ }
        \If{ \textcolor{black} {$N_{neg} > 0$} }
        \State \textcolor{black} {go to line 16 of this algorithm}
        \Else 
        \State \Return $G_{m}$ obtained using Eq. (1)
        \EndIf 
		\ElsIf{$N_{neg} > T$}
		\State $Low \leftarrow s_i$
		\Else
		\State $High \leftarrow s_i$
		\EndIf
		\State $i \leftarrow i + 1$
		\EndWhile
	\end{algorithmic}
\end{algorithm}

\subsection{Variational Quantum Search and Binary Search} 

In the pursuit of the global minimum, we employ an iterative approach that combines binary search and VQS. In each iteration, the VQS is applied to search for the negative values of the states after the adder, represented as $|\phi_2(s_i)\rangle$ and shown in Figure \ref{fig:vqs_binary}. Subsequently, the number of negative values ($N_{neg}$) is used as a key metric for establishing the termination criteria of the binary search and for modifying its lower and upper limits. 
The binary search can help to quickly shift the position of the function, as illustrated in Figure \ref{fig:high-level-graph}, such that the $N_{neg}$ for the shifted function is between 1 and $T$, usually within several iterations. 


\subsubsection{Variational Quantum Search (VQS)}
\label{subsec:vqs}

VQS is a variational quantum algorithm (VQA) proposed in \cite{zhan2023variational}. As a variational alternative to Grover's algorithm, VQS has undergone testing on searching unstructured databases comprising up to 26 qubits. The goal of VQS is to search good elements in an unstructured database, where the good elements' label qubit in the VQS circuit is in state $|1\rangle$. VQS employs an Ansatz $U(\theta)$ from \cite{zhan2023variational}, which is optimized to transform the quantum state from a superposition of all elements to a superposition of solely good elements. The objective function in VQS is expressed as:
\begin{equation}
	f_{\text{obj}}(\theta) = -0.5(\langle Z_1\rangle - \langle Z_2\rangle)
 \label{eq:obj}
\end{equation}
where $\langle Z_1\rangle$ and $\langle Z_2\rangle$ are measurements of two quantum circuits given in Figure \ref{fig:vqs_binary}, achieved by the Hadamard test 
\cite{hadamardtest}, and can be expressed as Eqs. (\ref{eq:Z_1}) and (\ref{eq:Z_2}):
\begin{equation}
    \langle Z_1\rangle = \langle\phi_2(s_i)|U(\theta)|\phi_2(s_i)\rangle
    \label{eq:Z_1}
\end{equation}
\begin{equation}
    \langle Z_2\rangle = \langle\phi_2(s_i)|Z\otimes I^{\otimes n} U(\theta)|\phi_2(s_i)\rangle\text{.}
    \label{eq:Z_2}
\end{equation}

The MSQ of $|\phi_2(s_i)\rangle$, given in Eq. (\ref{eq:adder_2}), corresponds to the overflow qubit described in Section \ref{sec:prereq}, and serves as the label qubit of VQS. 
When $U(\theta)$ is applied to this state, the resulting state is shown in Eq. (\ref{eq:vqs_res}).
\begin{equation}
	|\phi_3(s_i, \theta)\rangle = U(\theta) |\phi_2(s_i)\rangle
	\label{eq:vqs_res}
\end{equation}

\begin{figure}[!h]
    \centering
    \begin{subfigure}[b]{\textwidth}
        \centering
        \scalebox{1}{
            \begin{tikzpicture}[label distance=4mm]
                \coordinate (reference) at (8.2,-3);
                \coordinate (reference2) at (7.5,-3);

                \draw[rounded corners, dashed, style={darkgreen, thick, shorten <= -3mm, shorten >= -3mm}] ($(reference) + (-6.6,0.7)$) rectangle ($(reference) + (-3.7,2.3)$) node[midway, above, yshift=8mm, text=darkgreen]{Sign Extension};

                \draw[rounded corners, solid, style={blue, thick, shorten <= -3mm, shorten >= -3mm}] ($(reference) + (-0.3,-1.4)$) rectangle ($(reference) + (5,3.4)$) node[midway, above, yshift=25mm, text=blue]{VQS};

                \begin{yquant}[operators/every barrier/.append style={red, thick, shorten <= -3mm, shorten >= -3mm}, register/separation=3mm]

                    qubit {} ht;
                    qubit {} o; 

                    qubit {$\ket{0}_{n+m-1}$} ya;
                    qubit {$\ket{0}_{n+m-2}$} yb;
                    nobit ydots;
                    qubit {$\ket{0}_{n}$} yc;

                    qubit {$\ket{0}_{n-1}$} xa;
                    qubit {$\ket{0}_{n-2}$} xb;
                    nobit xdots;
                    qubit {$\ket{0}_{0}$} xc;

                    discard ht;
                    discard o;

                    hspace {2mm} ht;
                    hspace {2mm} o;
                    hspace {2mm} ya;
                    hspace {2mm} yb;
                    hspace {2mm} ydots;
                    hspace {2mm} yc;
                    hspace {2mm} xa;
                    hspace {2mm} xb;
                    hspace {2mm} xdots;
                    hspace {2mm} xc;

                    h xa;
                    h xb;
                    text {$\rvdots$} xdots;
                    text {$\rvdots$} ydots;
                    h xc;

                    ["$|\phi_0\rangle$"]
                    barrier (-);
                    init {$overflow\ket{0}$} o;
                    box {$\ \hat{O}_f \ $} (xa, xb, xdots, xc, ya, yb, ydots, yc);

                    text {$MSQ$} ya;
                    cnot o | ya; 

                    hspace {5mm} xa;
                    hspace {5mm} xb;
                    hspace {5mm} xc;

                    ["$|\phi_1\rangle$"]
                    barrier (ht,o,ya, yb, ydots, yc);


                    discard xa;
                    discard xb;
                    discard xdots;
                    discard xc;

                    box {$\ A(s_i)\ $} (o,ya, yb, ydots, yc);

                    ["$|\phi_2(s_i)\rangle$"]
                    barrier (ht,o,ya, yb, ydots, yc);

                    init {$\ket{0}$} ht;
                    hspace {2mm} ht;
                    h ht;
                    box {$\ U(\theta) \ $} (o,ya, yb, ydots, yc) | ht; 
                    h ht;
                    hspace {2mm} ht;
                    text {$\langle Z_1\rangle$} ht;
                    discard ht;

                \end{yquant}
            \end{tikzpicture}}
        \caption{}
        \label{fig:vqs_binary_a}
    \end{subfigure}

    \begin{subfigure}[b]{\textwidth}
        \centering
        \scalebox{1}{
            \begin{tikzpicture}[label distance=4mm]
                \coordinate (reference) at (8.2,-3);
                \coordinate (reference2) at (8.5,-3);

                \draw[rounded corners, dashed, style={darkgreen, thick, shorten <= -3mm, shorten >= -3mm}] ($(reference) + (-6.6,0.6)$) rectangle ($(reference) + (-3.7,2.2)$) node[midway, above, yshift=8mm, text=darkgreen]{Sign Extension};

                \draw[rounded corners, solid, style={blue, thick, shorten <= -3mm, shorten >= -3mm}] ($(reference) + (-0.50,-1.6)$) rectangle ($(reference) + (5.4,3.2)$) node[midway, above, yshift=24mm, text=blue]{VQS};

                \begin{yquant}[operators/every barrier/.append style={red, thick, shorten <= -3mm, shorten >= -3mm}, register/separation=3mm]

                    qubit {} ht;
                    qubit {} o;

                    qubit {$\ket{0}_{n+m-1}$} ya;
                    qubit {$\ket{0}_{n+m-2}$} yb;
                    nobit ydots;
                    qubit {$\ket{0}_{n}$} yc;

                    qubit {$\ket{0}_{n-1}$} xa;
                    qubit {$\ket{0}_{n-2}$} xb;
                    nobit xdots;
                    qubit {$\ket{0}_{0}$} xc;

                    discard ht;
                    discard o;

                    hspace {2mm} ht;
                    hspace {2mm} o;
                    hspace {2mm} ya;
                    hspace {2mm} yb;
                    hspace {2mm} ydots;
                    hspace {2mm} yc;
                    hspace {2mm} xa;
                    hspace {2mm} xb;
                    hspace {2mm} xdots;
                    hspace {2mm} xc;

                    h xa;
                    h xb;
                    text {$\rvdots$} xdots;
                    text {$\rvdots$} ydots;
                    h xc;

                    ["$|\phi_0\rangle$"]
                    barrier (-);
                    init {$overflow\ket{0}$} o;
                   
                    box {$\ \hat{O}_f \ $} (xa, xb, xdots, xc, ya, yb, ydots, yc);
                    text {$MSQ$} ya;
                    cnot o | ya; 

                    hspace {5mm} xa;
                    hspace {5mm} xb;
                    hspace {5mm} xc;

                    ["$|\phi_1\rangle$"]
                    barrier (ht,o,ya, yb, ydots, yc);


                    discard xa;
                    discard xb;
                    discard xdots;
                    discard xc;
                    
                    box {$\ A(s_i)\ $} (o,ya, yb, ydots, yc);

                    ["$|\phi_2(s_i)\rangle$"]
                    barrier (ht,o,ya, yb, ydots, yc);

                    init {$\ket{0}$} ht;
                    hspace {2mm} ht;
                    h ht;
                    box {$\ U(\theta) \ $} (o,ya, yb, ydots, yc) | ht; 

                    hspace {2mm} o;
                    z o | ht; 
                    h ht;
                    hspace {2mm} ht;
                    text {$\langle Z_2\rangle$} ht;
                    discard ht;

                \end{yquant}
            \end{tikzpicture}
        }
        \caption{}
        \label{fig:vqs_binary2_b}
    \end{subfigure}
\caption{Two quantum circuits used for the QGMF. (a) and (b) depict quantum circuits employed to measure $\langle Z_1\rangle$ and $\langle Z_2\rangle$, which are utilized in Eq. (\ref{eq:obj}). \rev{Note that the VQS part does not have an oracle.}}
    \label{fig:vqs_binary}
\end{figure}




When the objective function specified in Eq. (\ref{eq:obj}) reaches its global minimum, the parameters of the Ansatz also reach their optimal values, i.e., $\theta = \theta_{\text{best}}$, terminating the VQS iteration \rev { \cite{zhan2023variational} }. 
At this point, the quantum state $|\phi_3(s_i, \theta_{\text{best}})\rangle$ represents a superposition solely of the good elements, or equivalently, only the negative values in the states after the adder. 
This state can also be expressed as Eq.~(\ref{eq:vqs_res2}), where $\beta_j$ represents the amplitude of the basis state $|y_j\rangle$, corresponding to the $j^{\text{th}}$ good element, or equivalently, the $j^{\text{th}}$ distinct negative value.
Then, the count of negative values, denoted by $N_{\text{neg}}$, is determined by measuring the state $|\phi_3(s_i, \theta_{\text{best}})\rangle$, i.e., this count is equal to the number of unique states measured. 

\begin{equation}
	\label{eq:vqs_res2}
	|\phi_3(s_i,\theta_{\text{best}})\rangle= \sum_{j=1}^{N_{neg}} \beta_j |y_j\rangle
\end{equation}

\rev{
Theoretically, Eq. (\ref{eq:vqs_res2}) includes only good elements. In other words, the $\beta_j$ corresponding to bad elements is 0. However, in practice, if a near-optimal $\theta$ is obtained instead of the global optimal, some bad elements may still be present.
}

\subsubsection{Binary Search}
We employ a binary search method, detailed in Algorithm~\ref{alg:binary_vqs}, to optimize the shift value $s_i$ iteratively, aiming to refine our search for the global minimum effectively. The algorithm initiates by setting the shift $s_i$ to the midpoint between the current lower and upper bounds, defined as $2^{n-1}$ and $-2^{n-1}$, respectively, where $n$ represents the number of qubits. 

In each iteration, we measure state $|\phi_3(s_i,\theta_{\text{best}})\rangle$ as specified in Eq. (\ref{eq:vqs_res2}) to determine the count of negative function values, denoted as $N_{neg}$. If the number of distinct negative values $N_{neg}^{(i)}$ exceeds a predefined threshold $T$, it suggests an excess of negative values, prompting an upward shift adjustment, and we assign the current $s_i$ to the lower bound of the subsequent iteration.
Conversely, if $N_{neg}^{(i)}$ is approximately zero, indicating that the shifted objective function values are predominantly above the x-axis, a downward shift is applied, and we assign the current $s_i$ to the upper bound of the subsequent iteration.

\revv{By checking lines 16, 22, and 27 of Algorithm 1, it is sufficient to determine whether $N_{neg}$ exceeds the predefined threshold $T$, without requiring the exact value of $N_{neg}$ if it is greater than $T$. During the measurement process, once the number of distinct negative values measured exceeds $T$, we terminate the measurement process for that iteration. Consequently, the maximum number of points we need to measure per iteration is limited to $T$, a small constant.}

The binary search continues until $N_{neg}^{(i)}$ falls within the range of 1 to $T$, at which point the process is terminated, and the minimum value among the detected negative values, $M_r$, is identified. The global minimum $G_m$ is then calculated using $G_m=M_r-s_i$, given in Eq. (\ref{eq:gmin}). Note that the threshold $T$ is a small constant, leading to a constant time complexity of $O(T)$ for searching through the remaining negative values.

\rev{
\subsubsection{Handle Unbalanced State Probabilities}
The probability of measuring any state $ |y_j\rangle $ is given by $ |\beta_j|^2 $. Let $ p_{\text{min}} $ denote the minimum probability among these states. To reliably measure all $ |y_j\rangle $ states, the number of shots should be large enough to ensure that even the least probable states are detected.

For ease of expression, we define the balanced and unbalanced coefficients below.

\textit{Balanced $ \beta_j $}: When the coefficients are balanced (i.e., none of the $ \beta_j $ are excessively small compared to others), Algorithm 1 performs effectively because all $ |y_j\rangle $ states are likely to be measured within a reasonable number of shots.

\textit{Unbalanced $ \beta_j $}: If some $ \beta_j $ coefficients are significantly smaller than others, the corresponding $ |y_j\rangle $ states have very low probabilities of being measured. This imbalance can lead to incomplete sampling of the solution space, potentially affecting the algorithm's performance.

To address the issue of unbalanced $\beta_j$, we integrate the following four steps into Algorithm 1.

Step 1 - Identifying Minimum $y_j$: When $N_{\text{neg}} \in [1, T]$, we begin by identifying the minimum value among all $y_j$, denoted as $y_{\text{min}}$.

Step 2 - Adjusting the Threshold $s_i$: We then increase the threshold $s_i$ by $|y_{\text{min}}| + \epsilon$, where $\epsilon$ is a very small positive value. This adjustment shifts all objective values upward, ensuring that the high-probability $\lvert y_j \rangle$ states (which were previously negative) become positive and are effectively filtered out.

Step 3 - Focusing on Low-Probability States: After the adjustment in Step 2, the VQS focuses on finding smaller objective values (i.e., values that are lower than $y_{\text{min}}$), if any. These smaller objective values were low-probability $\lvert y_j \rangle$ states that were not measured in Step 1. As all high-probability $\lvert y_j \rangle$ states were filtered out in the previous step, the probability of the states corresponding to these smaller objective values becomes high and they are much more likely to be measured, without the need to set a very high number of shots.

Step 4 - Iterative Measurement: If promising elements are detected among these low-probability states, we proceed with the QGMF method. Otherwise, we conclude that the $\beta_j$ values are balanced and terminate the algorithm.

These four steps are reflected in lines 17 - 23 of Algorithm 1. Note that the purpose of these four steps is to be able to measure low-probability states without setting the number of shots to an extremely large value in the case of unbalanced state probabilities. 

One of the advantages of using the QGMF approach is that it allows us to avoid setting an excessively large number of shots, even for a large number of qubits, $n$. Specifically, when $n$ is large, it can lead to numerous negative values ($ N_{neg}> T $). In such cases, the VQS can terminate the measurement process once more than $T$ distinct states have been measured, update the $Low$ value in Algorithm 1, and continue to the next iteration, as described in lines 27-28 of Algorithm 1. For this reason, we have chosen to set the number of shots to a fixed value, typically in the range of 5,000-10,000. 

}


\section{Complexity Analysis and Case Study}
\label{sec:tests}

This section delves into the complexity analysis and empirical simulation of the QGMF algorithm. Given that QGMF is a VQA, its performance metrics such as complexity and efficiency are influenced by several factors, including circuit depth, convergence rate, and number of measurements.

The adder circuit used in this paper consists of QFT and its inverse as well as one layer of Rz gates. Therefore, the complexity of the adder circuit, essential for shifting operations, is comparable to that of the QFT at time complexity of \(O(n^2))\), where $n$ is the number of qubits \cite{yuan2023improved}. The number of CR operations required for the QFT is $n(n+1)/2$. 
\rev{Furthermore, the VQS plays a pivotal role in the local search phase, featuring a linear depth complexity of $O(n)$, thereby reducing the required quantum resources.}

\rev{
In the QGMF method, we employ only one oracle whose circuit depth is crucial for understanding the computational complexity of our algorithm. The oracle, denoted as $ \hat{O}_f $ used in Eq. (\ref{eq:oracle1}), implements the function $ f(x) $ and prepares the quantum state $ |\phi_0'\rangle $. Preparing such an arbitrary quantum state is a well-studied problem in quantum computing. According to Theorem 1 in \cite{Zhang2022PRL}, any $ n $-qubit quantum state can be deterministically prepared using only single- and two-qubit gates with a circuit depth of $ \Theta(n) $ and $O(2^n)$ ancillary qubits. 

Note that in the VQS part of the circuit, as shown in Figure \ref{fig:vqs_binary}, the $U(\theta)$'s depth is $3(n+1)$ and 15 if using type-I and type-II Ansatzs, respectively, according to \cite{zhan2023variational}. In summary, the QGMF has a $O(n)$-depth circuit complexity, including all circuit parts such as the oracle, the QFT based adder, and the VQS. 


}

The optimization complexity, critical for identifying the optimal parameters of VQS, can be quantified based on the number of iterations and measurements.  According to the findings documented in \cite{zhan2023variational,zhan2023nearperfect}, less than 300 iterations are necessary to achieve the desired state in most runs, highlighting the algorithm's efficiency in parameter optimization. The number of measurements is twice the number of iterations. 

Recent studies have confirmed the effectiveness of VQS in systems with up to 26 qubits, as detailed in \cite{zhan2023variational,zhan2023nearperfect}. Efforts are currently underway to extend VQS capabilities to more qubits. For computational complexity analysis, it is important to recognize that in scenarios involving more than 26 qubits - where VQS's efficiency has not yet been confirmed - Grover's Search algorithm can be utilized as an interim solution. According to \cite{grover1996fast}, this algorithm offers a quadratic speedup. Despite this advantage, preliminary analysis suggests that VQS could potentially outperform Grover's algorithm in terms of circuit depth when applied to systems larger than 26 qubits. This suggests that the QGMF might achieve more than quadratic speedup in finding optimal solutions, indicating a promising avenue for further research in quantum computing efficiency.

\begin{figure}[H]
	\centering
	\includegraphics[scale=0.5]{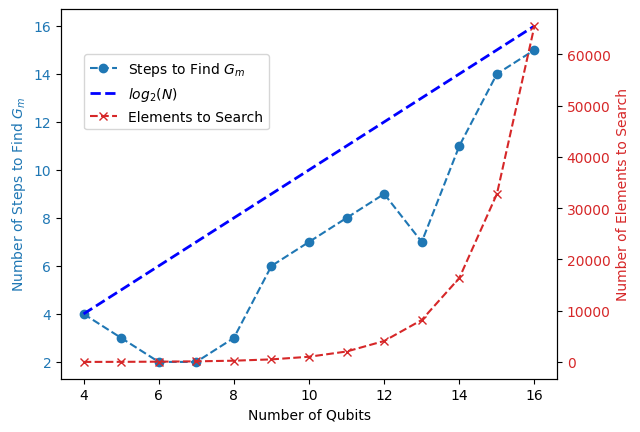}


\caption{
Efficiency of the QGMF in identifying the global minimum of a randomly generated oracle, indicated by the number of binary search steps required (left vertical axis) across various qubit counts. \rev{ The figure }contrasts this with the total number of elements a classical brute-force search must navigate to theoretically ensure pinpointing the global minimum (right vertical axis). The function $\log_2{(N)}$, \rev{where $N = 2^n$, is also depicted to illustrate the exponential growth of the search space. 
Note that while Appendix A discusses creating sparse quantum states with a small number of non-zero elements (calculated via the approach mentioned in Algorithm. \ref{alg:random_oracle_builder}), the dashed red  curve 
represents the worst-case scenario where all possible computational basis states are non-zero. This provides a clear comparison with classical search methods by illustrating the algorithm's efficiency even when dealing with the maximum possible number of non-zero elements.
}
}

	\label{res2}
\end{figure}

The outer loop of Algorithm 1 is a binary search, which is known to run in logarithmic time in the worst case. Therefore, we can expect the QGMF's outer iteration to converge quickly\rev{.} 
The results given in this figure are obtained 
from simulation using Pennylane 
\cite{bergholm2022pennylane, soltaninia2023comparison}. Simulation tests indicate that the QGMF is as accurate as the classical brute-force method. 
As shown in Figure \ref{res2}, the number of binary search steps required by the QGMF to locate the global minimum is within 15 for all the 13 cases we studied.
These simulations assessed QGMF's adaptability to different quantum oracles by encoding potential function values in quantum states. 
Each state involved a randomly generated, normalized input vector of dimension $2^n$, adhering to quantum circuitry norms. Detailed construction of the random oracle and comparisons with classical brute-force results are provided in Algorithm \ref{alg:random_oracle_builder}, underscoring the framework's potential efficacy in quantum computing applications.

\rev{
Note that the results in this section are based on perfect state vector calculations and we utilized the Pennylane simulation platform \cite{bergholm2022pennylane}.  

}

\rev{
\subsection{Apply QGMF to Determine the Chromatic Number of a Graph}

The \textit{chromatic number} of a graph, typically denoted as $\chi(G)$, is one of the most fundamental and widely studied concepts in graph theory. It represents the smallest number of distinct colors required to color the vertices of a graph $G$ such that no two adjacent vertices share the same color. This associated problem, known as the \textit{vertex coloring problem}, arises in numerous real-world applications, from scheduling and resource allocation to frequency assignment in wireless networks and register allocation in compilers.

Despite its seemingly simple formulation, determining the chromatic number is a highly complex task. The problem is classified as \textit{NP-hard}, meaning that no polynomial-time algorithm is known for finding the chromatic number for arbitrary graphs. This makes the chromatic number a critical subject of study not only in theoretical computer science but also in practical fields where optimization and resource minimization are crucial.

The combinatorial nature of the problem means that as the size and complexity of the graph grow, finding the chromatic number becomes increasingly difficult. This is where QGMF offers a promising alternative to classical methods. 

\subsubsection{Problem Definition}
The vertex coloring problem is a classic combinatorial optimization problem that seeks to assign colors to the vertices of a graph \( G = (V, E) \), where \( V \) is the set of vertices and \( E \) is the set of edges, such that no two adjacent vertices share the same color. 
Formally, our objective is to find a function \( f : V \to \{1, 2, \dots, k\} \) that assigns a color \( f(v) \) to each vertex \( v \in V \), where \( k \) represents the number of available colors, subject to the constraint that for every edge \( (u, v) \in E \), \( f(u) \neq f(v) \). 

The goal is to determine the minimum number of colors \( k \) required to achieve a proper coloring, known as the \textit{chromatic number} \( \chi(G) \), defined as:
\[ \chi(G) = \min \{k : \text{there exists a valid coloring with } k \text{ colors}\} \]

 As an illustration, in Figure \ref{graph}, we have a graph with four vertices \( V_1, V_2, V_3, V_4 \), where the edges represent adjacency constraints. The edges between the vertices are defined as follows: \( (V_1, V_2) \), \( (V_1, V_4) \), \( (V_2, V_3) \), \( (V_2, V_4) \), and \( (V_3, V_4) \). The objective is to assign a color to each vertex such that no two adjacent vertices share the same color.
In this case, the constraints are:
\[
f(V_1) \neq f(V_2), \quad f(V_1) \neq f(V_4), \quad f(V_2) \neq f(V_3), \quad f(V_2) \neq f(V_4), \quad f(V_3) \neq f(V_4)
\]
where \( f(V_i) \) denotes the color assigned to vertex \( V_i \).

\begin{figure}[h]
    \centering
    \includegraphics[width=0.3\textwidth]{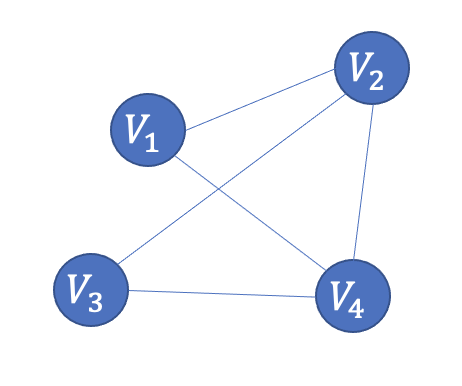}
    \caption{An example graph with four vertices.}
    \label{graph}
\end{figure}


\subsubsection{Binary Encoding of Colors}
We encode the colors assigned to each vertex using binary variables. For a graph with \( n \) vertices and up to \( k \) colors, the colors can be represented using \( \lceil \log_2 k \rceil \) qubits. Therefore, we need \( n \times \lceil \log_2 k \rceil \) qubits, to encode the colors of the vertices. We refer to this encoding as \textbf{Color Encoding}.

In Figure \ref{graph}, for example, we have a graph with four vertices \( V_1, V_2, V_3, V_4 \), with edges connecting \( (V_1, V_2) \), \( (V_1, V_4) \), \( (V_2, V_3) \), \( (V_2, V_4) \), and \( (V_3, V_4) \). To determine the chromatic number, we encode the color assignments into qubits. Since we want to use up to four different colors, we assign two qubits per vertex to represent the binary encoding of the colors. For instance, the colors can be encoded as follows:
\[
|00\rangle \rightarrow \text{Red}, \quad |01\rangle \rightarrow \text{Green}, \quad |10\rangle \rightarrow \text{Blue}, \quad |11\rangle \rightarrow \text{Yellow}
\]


\subsubsection{Constraint Encoding}
To enforce the constraints, we use a \textit{controlled-U} gate \cite{cu}, where the quantum adder \( A(1) \) (refer to Section \ref{sec:tests}) is applied as the operation \( U \). This allows us to build a controlled quantum adder, denoted as \( \textit{controlled-A}(1) \), as shown in Figure \ref{fig:ctrl_a}. 

\begin{figure}[!h]
\centering
\begin{quantikz}
&\ctrl{1} &\qw &\\
&\gate[3][1.7cm]{A(1)} &\qw & \\
& & \qw& \\
& & \qw& 
\end{quantikz}

\caption{Controlled Quantum Adder circuit \( \textit{controlled-A}(1) \).}

\label{fig:ctrl_a}
\end{figure}

For each constraint, which involves two adjacent nodes in the graph, we apply a \( \textit{controlled-A}(1) \). If two adjacent vertices share the same color (i.e., their qubit states are the same), the adder circuit adds \( |1\rangle \) to the \textbf{Counter Register (C)}, which tracks the number of violations and stores this count in binary. In the worst-case scenario, where all vertices of a fully connected graph are assigned the same color, the number of violations reaches its maximum. 

As an example, for the graph shown in Figure \ref{graph}, the corresponding quantum circuit is constructed as illustrated in Figure \ref{fig:vc_example} for a single color state. The complete circuit requires extensions to account for all possible color states. 

\begin{figure}[!h]
\centering
\begin{quantikz}[row sep=0.5cm]
\lstick{$V_{10}$} & \qw & \ctrl{10}              & \qw & \ctrl{9} & \qw & \qw  & \qw & \qw & \qw  & \qw & \qw & \qw \\
\lstick{$V_{11}$} & \qw & \ctrl{9}             & \qw & \ctrl{8} & \qw & \qw  & \qw  & \qw & \qw  & \qw& \qw & \qw \\
\lstick{$V_{20}$} & \qw & \ctrl{8}            & \qw & \qw & \qw &  \ctrl{7} & \qw & \ctrl{7}& \qw  & \qw& \qw & \qw  \\
\lstick{$V_{21}$} & \qw & \ctrl{7}           & \qw& \qw& \qw & \ctrl{6}  & \qw &\ctrl{6}& \qw  & \qw& \qw & \qw  \\
\lstick{$V_{30}$} & \qw & \qw                  & \qw & \qw  & \qw & \ctrl{5}  & \qw & \qw & \qw & \ctrl{4} & \qw& \qw \\
\lstick{$V_{31}$} & \qw  & \qw                 & \qw & \qw & \qw & \ctrl{4}  & \qw & \qw & \qw   & \ctrl{3} & \qw& \qw \\
\lstick{$V_{40}$} & \qw & \qw                 & \qw& \ctrl{2} & \qw & \qw & \qw  & \ctrl{2}& \qw & \ctrl{2} & \qw & \qw  \\
\lstick{$V_{41}$} & \qw & \qw                  & \qw & \ctrl{1} & \qw & \qw & \qw  &\ctrl{1}& \qw &\ctrl{1}  & \qw& \qw  \\
\lstick[5]{Counter Register}&\lstick{$\ket{0}$} & \gate[wires=5]{A(1)} & \qw & \gate[wires=5]{A(1)}  & \qw & \gate[wires=5]{A(1)} & \qw  & \gate[wires=5]{A(1)} & \qw  & \gate[wires=5]{A(1)}  \qw & \qw  & \qw \\
\qw&\lstick{$\ket{0}$} & \qw & \qw  & \qw & \qw & \qw  & \qw  & \qw  & \qw  & \qw & \qw  & \qw \\
\qw&\lstick{$\ket{0}$} & \qw & \qw  & \qw & \qw & \qw  & \qw  & \qw  & \qw  & \qw & \qw  & \qw \\
\qw&\lstick{$\ket{0}$} & \qw & \qw  & \qw & \qw & \qw  & \qw  & \qw  & \qw  & \qw & \qw  & \qw \\ 
\qw&\lstick{$\ket{0}$} & \qw & \qw & \qw & \qw & \qw  & \qw  & \qw  & \qw  & \qw & \qw  & \qw
\end{quantikz}
\caption{Binary encoding of vertex colors and color assignment constraints for the color state \(|11\rangle\) for the graph shown in Figure \ref{graph}, which has four vertices. The full circuit includes three additional copies of this structure, covering all four possible colors. Each vertex is represented by two qubits, encoding one of the four possible colors.}

\label{fig:vc_example}
\end{figure}

\subsubsection{Efficient Search to Find \( \chi(G) \)}

Using the counter register as the oracle for the QGMF, we can determine the minimum number of violations based on our color inputs. To find the chromatic number \( \chi(G) \), we employ a binary search over the possible number of colors. 
By using binary search, we aim to find the smallest possible \( \chi(G) \), with the binary search terminating when the number of violations reaches a near-zero probability and remains stable. At that point, we can obtain a good estimate of \( \chi(G) \).

For example, in Figure \ref{graph}, we need at least one color and at most four colors (Figure \ref{fig:vc_colors}a, b and c show the binary encoding for four, two, and one colors, respectively). We start by considering the middle integer, which is two. To encode two colors for each vertex, we can apply a Hadamard gate on one of the two qubits used to encode the vertex colors, as shown in Figure \ref{fig:vc_colors}b. The number of violations is $2$, so we update our lower bound to $2$ and select the middle integer between $2$ and $4$, which is $3$. As shown in Figure \ref{fig:min_violations}, this satisfies our termination condition, and we determine that the chromatic number \( \chi(G) \) is $3$.

\begin{figure}[!h]
\centering

\begin{subfigure}[b]{0.33\textwidth}
\centering
\begin{quantikz}
\lstick{$\ket{0}$} & \gate{H} & \qw & \rstick{$V_{10}$}  \\
\lstick{$\ket{0}$} & \gate{H}& \qw & \rstick{$V_{11}$} 
\end{quantikz}
\caption{$n_c=4$}
\end{subfigure}
\hfill
\begin{subfigure}[b]{0.33\textwidth}
\centering
\begin{quantikz}
\lstick{$\ket{0}$} & \qw& \qw & \rstick{$V_{10}$}  \\
\lstick{$\ket{0}$} &\gate{H}& \qw & \rstick{$V_{11}$} 
\end{quantikz}
\caption{$n_c=2$}
\end{subfigure}
\hfill
\begin{subfigure}[b]{0.33\textwidth}
\centering
\begin{quantikz}
\lstick{$\ket{0}$} & \qw & \qw & \rstick{$V_{10}$}  \\\\
\lstick{$\ket{0}$} & \qw& \qw & \rstick{$V_{11}$} 
\end{quantikz}
\caption{$n_c=1$}
\end{subfigure}

\caption{Binary encoding of different number of colors for vertex $V_1$ of graph in Figure \ref{graph}
, where \(|00\rangle\), \(|01\rangle\), \(|10\rangle\), and \(|11\rangle\) represent red, green, blue, and yellow, respectively.}
\label{fig:vc_colors}
\end{figure}

\begin{figure}[h]
\centering
\begin{tikzpicture}
\begin{axis}[
    xlabel={Number of Colors \( (k) \)},
    ylabel={Minimum Number of Violations},
    xtick={1,2,3,4},
    ytick={0,1,2,3,4},
    ymin=0, ymax=5,
    xmin=0.5, xmax=4.5,
    grid=both,
    ymajorgrids=true,
    xmajorgrids=true,
    width=0.7\textwidth,
    height=0.5\textwidth,
    tick label style={font=\small},
    label style={font=\small},
]

\addplot[
    only marks,
    mark=*,
    mark size=3pt,
] coordinates {
    (1,4)
    (2,2)
    (3,0)
    (4,0)
};

\addplot[
    only marks,
    mark=*,
    mark size=3pt,    
    color=darkgreen,      
] coordinates {
    (3,0)
};

\node at (axis cs:3,0.2) [anchor=south] { \textbf{\( \chi(G) \)}};

\end{axis}
\end{tikzpicture}
\caption{Graph of the minimum number of violations versus the number of colors \( k \) for graph shown in Figure \ref{graph}.}
\label{fig:min_violations}
\end{figure}

}
\section{Discussion}
\label{sec:conc}
In this paper, we introduce the QGMF, an innovative approach designed to efficiently locate the global minimum of quantum oracles. The QGMF integrates a binary search with the VQS, harnessing quantum computing's potential to outperform traditional brute-force methods significantly. Our analysis confirms the effectiveness of QGMF, demonstrating its superior performance in complex optimization scenarios. 
The implications of our study are profound, opening new avenues in fields such as finance, chemistry, and machine learning. 

Future research will focus on evaluating QGMF under realistic quantum conditions, expanding its applications across various sectors, and refining quantum oracles to enhance their utility and performance. Despite its promising outcomes, we recognize the limitations of our current study, primarily the assumption of ideal quantum computing conditions which might not accurately represent the complexities faced in noisy quantum environments. Additionally, the scalability and practical deployment of QGMF in real-world applications require further exploration to fully realize its potential.

\newpage 

\begin{appendices}
\section{\revv{Data Embedding}}
\label{sec:prereq}

This section outlines data embedding techniques in quantum computing, focusing on basis embedding and 2's complement representation. It highlights the advantages of 2's complement in simplifying arithmetic operations within quantum systems, providing a foundation for its application in the QGMF, especially detailed in Section \ref{subsec:shifting}. Furthermore, the 2's complement described in this section provides a label qubit that is an essential input to VQS, a core part of QGMF, as described in Section \ref{subsec:vqs}.



\begin{table}[h]
	\centering
	\begin{tabular}{|c|c|c|c|}
		\hline
		Binary & Signed Magnitude & 1's complement & 2's complement \\ \hline
		$|0000\rangle$ &\ \ \textbf{\textcolor{darkred}{0}} &\ \ \textbf{\textcolor{darkred}{0}} &\ \ \textbf{\textcolor{darkgreen}{0}}\\ \hline
		$|0001\rangle$ &\ \ 1 &\ \ 1 & \ \ 1 \\ \hline
		$|0010\rangle$ &\ \ 2 &\ \ 2 &\ \ 2 \\ \hline
		$|0011\rangle$ &\ \ 3 &\ \ 3 &\ \ 3 \\ \hline
		$|0100\rangle$ &\ \ 4 &\ \ 4 &\ \ 4 \\ \hline
		$|0101\rangle$ &\ \ 5 &\ \ 5 &\ \ 5 \\ \hline
		$|0110\rangle$ &\ \ 6 &\ \ 6 &\ \ 6 \\ \hline
		$|0111\rangle$ &\ \ 7 &\ \ 7 &\ \ 7 \\ \hline
		$|1000\rangle$ &\ \ \textbf{\textcolor{darkred}{0}} &$-7$ &$-8$ \\ \hline
		$|1001\rangle$ & $-1$ & $-6$ & $-7$\\ \hline
		$|1010\rangle$ & $-2$ & $-5$ & $-6$ \\ \hline
		$|1011\rangle$ & $-3$ & $-4$ & $-5$ \\ \hline
		$|1100\rangle$ & $-4$ & $-3$ & $-4$ \\ \hline
		$|1101\rangle$ & $-5$ & $-2$ & $-3$ \\ \hline
		$|1110\rangle$ & $-6$ & $-1$ & $-2$ \\ \hline
		$|1111\rangle$ & $-7$ &\ \ \textbf{\textcolor{darkred}{0}} & $-1$ \\ \hline
	\end{tabular}
	\caption{Signed binary representations for 4-qubit encoding. It is evident that Signed Magnitude and 1's complement have dual representations for the number $0$, which proves inefficient in arithmetic operations. The 2's complement emerges as the standard method for encoding signed integer numbers in binary format \cite{sign_binary}.}
	\label{sbinary}
\end{table}

\subsection{Basis Embedding}
In quantum computing, there are several ways to encode data like basis embedding, angle embedding, amplitude embedding, and so on \cite{bergholm2022pennylane}. In this paper, we used basis embedding.
Basis embedding associates each input with a computational basis state of a qubit system. Therefore, data has to be in the form of binary strings. The embedded quantum state is the bit-wise translation of a binary string to the corresponding states of the quantum subsystems. For example, \(x = 0011\) is represented by the 4-qubit quantum state \(|0011\rangle\).

\subsection{2's Complement Representation}
Various methods exist for encoding signed integer numbers in binary representation, including Signed Magnitude, 1's complement, and 2's complement \cite{sign_binary}. In Table~\ref{sbinary}, different representations of signed integer numbers encoded in a 4-qubit binary format are shown.
2's complement serves as a fundamental method for representing signed integers in binary form and its format also ensures that the number $0$ has only one representation, making it unambiguous \cite{sign_binary}. In this representation, the leftmost digit serves as the sign indicator: "0" signifies a positive number, and "1" denotes a negative number. Additionally, 
$p$ binary digits can accommodate a range of integer values ($R_p$) as defined in Eq. (\ref{eq:range_sign}).

\begin{equation}
	\label{eq:range_sign}
        R_p = [-2^{p-1},2^{p-1}-1]
\end{equation}

The concepts discussed are also applicable to the basis embedding of a 2's complement quantum binary state $|x\rangle$, where the Most Significant Qubit (MSQ) signifies the sign of $|x\rangle$, in accordance with Eq. (\ref{eq:sign}).

\begin{equation}
	\label{eq:sign}
	sign(|x\rangle)=|x\rangle_{MSQ}= 
	\begin{cases}
		|0\rangle, &  x \geq 0 \\
		|1\rangle, &  x < 0
	\end{cases}
\end{equation}


\paragraph{Positive Numbers:}
Converting a positive decimal integer $x$ to 2's complement is straightforward, resembling the process for unsigned binary representation \cite{sign_binary}. For example $\text{Decimal 5} \rightarrow |\text{0101}\rangle$.


\paragraph{Negative Numbers:}
To convert negative decimal integer $x$ to 2's complement, we use the procedure below:

\begin{enumerate}
	\item Obtain the binary representation of the corresponding positive number ($-x$).
	\item Invert all qubits (change 0s to 1s and vice versa).
	\item Add 1 to the result.
\end{enumerate}
For example, $\text{Decimal $-$5} \rightarrow |\text{$\overline{0101}$}	+ 1\rangle = |1011\rangle$. 
\paragraph{Addition and Subtraction:} In the 2's complement, addition and subtraction operations mirror those of their unsigned counterparts, with a crucial consideration for the number of qubits required to accommodate the result. For instance, adding two 4-qubit numbers in 2's complement demonstrates this:

\[
\begin{array}{ccccc}
	& &\ |1010\rangle & \text{($-6$ in 4-qubit 2's complement)} \\
	& + &\ |1101\rangle & \text{($-3$ in 4-qubit 2's complement)} \\
	\hline
	& &\ |0111\rangle & \text{(Incorrect positive result)}
\end{array} 
\]

The mistaken positive result from adding two negative numbers highlights the inadequacy of 4 qubits for representing $-9$, which necessitates at least 5 qubits according to Eq. (\ref{eq:range_sign}). By expanding the operands to 5 qubits we have:

\[
\begin{array}{ccccc}
	& &\ |11010\rangle & \text{($-6$ extended to 5-qubit 2's complement)} \\
	& + &\ |11101\rangle & \text{($-3$ extended to 5-qubit 2's complement)} \\
	\hline
	& &\ |10111\rangle & \text{($-9$ in 5-qubit 2's complement)}
\end{array} 
\]

This adjustment accurately preserves the negative outcome, and the introduced qubit is termed an \textit{overflow} qubit. Notably, in 2's complement representation, extending a number necessitates the preservation of its sign, achieved by duplicating the MSQ before appending it, a rule named Sign Extension \cite{sign_binary}.

Thus, effectively conducting addition and subtraction in 2's complement entails:

\begin{enumerate}
	\item Converting all numbers to 2's complement, ensuring uniform qubit length.
	\item Avoiding overflow by adding necessary overflow qubits, 
 e.g., adding two 4-qubit numbers may require a 5th qubit for accurate representation. Therefore, we need to have at least 5 qubits.
    \item Finally, carry out addition or subtraction using these extended representations. Note that in this context, subtraction, expressed as "$a - b$", can be performed by adding "$a$" to the 2's complement of "$b$".
\end{enumerate}

\paragraph{Multiplication:} Suppose we have two numbers, $m$ and $k$, represented in unsigned binary format. We can express these numbers as
$
m = \sum_{i=0}^{n_m-1} m_i 2^i
$
and
$
k = \sum_{j=0}^{n_k-1} k_j 2^j
$, where, $m_i$ and $k_j$ are the individual qubits of $m$ and $k$, taking values from the set $\{0, 1\}$, and $n_m$ and $n_k$ represent the number of qubits used to represent $m$ and $k$, respectively. The multiplication of $m$ and $k$ can be expressed as $m \cdot k = \sum_{i=0}^{n_m-1} \sum_{j=0}^{n_k-1} m_i k_j 2^{i+j}
$. To perform this multiplication in a way compatible with 2's complement representation, we can follow these steps:
\begin{enumerate}
	\item First, convert both $m$ and $k$ into their respective 2's complement representations.	
	\item Next, expand the representation of both numbers by adding overflow qubits to make them each consist of $n_m + n_k$ qubits. Ensure that the added overflow qubits replicate the MSQ according to the sign extension rule. 
	\item After extending the representations, carry out the standard binary multiplication.
\end{enumerate}

For example, consider a scenario where we have a 3-qubit binary number $m$ and a 4-qubit binary number $k$, and we wish to calculate the product $m \times k$. If $m=-2$ and $k=-3$, their binary representations would be $m = |110\rangle$ and $k = |1101\rangle$.
Utilizing the sign extension rule and incorporating overflow qubits extends both numbers to 7 qubits. Thus, we get $m=|1111110\rangle$ and $k=|1111101\rangle$. In unsigned binary, these extended versions of $m$ and $k$ denote 126 and 125, respectively. When these numbers are multiplied in their unsigned form, the result is $m \times k = 15750$. This product can be represented in a 7-qubit binary format as $|0000110\rangle$, which corresponds to 6 in 2's complement notation.  


\rev{
\section{Random Oracle Builder Function}\label{secA1}
In this section, we present the pseudocode for the \emph{Random Oracle Builder} function (Algorithm~\ref{alg:random_oracle_builder}), which generates test vectors and their associated global minimum values to analyze our proposed algorithm.

Algorithm~\ref{alg:random_oracle_builder} generates a random, sparse vector \texttt{rand\_vec}, where each non-zero element represents the amplitude of a computational basis state in the quantum superposition. These amplitudes are assigned randomly to simulate complex and challenging optimization problems. Unlike the equal superposition produced by applying an oracle $O_f$, where each computational basis state has a probability of $2^{-n}$, our approach may produce probabilities less than $2^{-n}$. This design intentionally introduces additional complexity to demonstrate the robustness and efficiency of the QGMF algorithm under more difficult conditions.  

The vector \texttt{rand\_vec} specifically represents the amplitudes assigned to non-zero computational basis states. Non-zero elements in the vector correspond to the possible values generated by the function (oracle), while zero elements indicate outcomes not produced. By creating a sparse quantum state with non-zero amplitudes assigned to only a subset of basis states, we increase the variability and complexity of the optimization problem. The sparsity of \texttt{rand\_vec} is determined by selecting a random number of non-zero elements, generated using the following code snippet: num\_ones = numpy.random.randint(1, $\lfloor$ size/100$\rfloor$ + 2)

Here, \texttt{size} is $2^n$, which represents the total number of possible computational basis states for $n$ qubits. This approach ensures that the quantum state remains sparse by including only a small proportion (approximately 1\%) of non-zero elements, making the optimization task more challenging and realistic.



\begin{algorithm}
    \caption{Random Oracle Builder}\label{alg:random_oracle_builder}
    \begin{algorithmic}[1]
        \State \textbf{Input:} $n$: Number of qubits
        \State \textbf{Output:} \texttt{rand\_vec}: Normalized random vector, $G_{m}$: Global minimum
        \State
        \State $N \gets 2^{n}$
        \State $negative\_index \gets \lfloor N / 2 \rfloor$
        \State Initialize \texttt{rand\_vec} as a zero vector of length $N$
        \State $classical\_min\_index \gets N + 1$
        \State $min\_index\_found \gets$ False
        \rev{ \State Set the number of non-zero elements $\#\mathrm{non\_zeros} \leftarrow \operatorname{RandomInteger}\left(1, \left\lfloor N/100 \right\rfloor + 2\right)$}
        \State $indices \gets$ randomly selected sorted indices from $N$, of length $\#\mathrm{non\_zeros}$
        \For{each $i$ in $indices$}
        \State Assign a random value to \texttt{rand\_vec}$[i]$
        \If{not $min\_index\_found$}
        \If{$i < negative\_index$ and $i < classical\_min\_index$}
        \State $classical\_min\_index \gets i$
        \ElsIf{$i \geq negative\_index$}
        \State $classical\_min\_index \gets i$
        \State $min\_index\_found \gets$ True
        \EndIf
        \EndIf
        \EndFor
        \State Normalize \texttt{rand\_vec}
        \If{$classical\_min\_index < negative\_index$}
        \State $G_{m} \gets classical\_min\_index$
        \Else
        \State $sign\_bin\_rep \gets$ 2's complement of $classical\_min\_index$
        \State $G_{m} \gets$ convert $sign\_bin\_rep$ to integer
        \EndIf
        \State \Return{\texttt{rand\_vec}, $G_{m}$}
    \end{algorithmic}
\end{algorithm}

}




\end{appendices}


\section*{Code Availability} 

The code supporting this study is available on GitHub at the following URL:\\ \href{https://github.com/natanil-m/quantum_global_minimum_finder}{https://github.com/natanil-m/quantum\_global\_minimum\_finder}.

\section*{Data Availability} 

The datasets generated and/or analysed during the current study are available in the GitHub repository, \href{https://github.com/natanil-m/quantum_global_minimum_finder}{https://github.com/natanil-m/quantum\_global\_minimum\_finder}.

\section*{Acknowledgement} 
This research was supported by the NSF ERI program, under award number 2138702. This work used the Delta system at the National Center for Supercomputing Applications through allocation CIS220136 and CIS240211 from the Advanced Cyberinfrastructure Coordination Ecosystem: Services \& Support (ACCESS) program, which is supported by National Science Foundation grants \#2138259, \#2138286, \#2138307, \#2137603, and \#2138296. We acknowledge the use of IBM Quantum services for this work. The views expressed are those of the authors, and do not reflect the official policy or position of IBM or the IBM Quantum team.

\bibliography{sample}

\end{document}